\def\year{2021}\relax
\documentclass[letterpaper]{article} % DO NOT CHANGE THIS
\usepackage{aaai21}  % DO NOT CHANGE THIS
\usepackage{times}  % DO NOT CHANGE THIS
\usepackage{helvet} % DO NOT CHANGE THIS
\usepackage{courier}  % DO NOT CHANGE THIS
\usepackage[hyphens]{url}  % DO NOT CHANGE THIS
\usepackage{graphicx} % DO NOT CHANGE THIS
\usepackage{natbib}  % DO NOT CHANGE THIS AND DO NOT ADD ANY OPTIONS TO IT
\usepackage{caption} % DO NOT CHANGE THIS AND DO NOT ADD ANY OPTIONS TO IT
\usepackage{amsmath}
\usepackage{dirtytalk}
\title{An experiment on the mechanisms of racial bias in ML-based credit scoring in Brazil}
\author{
    Ramon Vilarino,\textsuperscript{\rm 1}\thanks{Author to whom all correspondence should be addressed}
    Renato Vicente\textsuperscript{\rm 2}\textsuperscript{\rm 3}
    \\
}
\affiliations{
    \textsuperscript{\rm 1}University of California, Berkeley\\
    \textsuperscript{\rm 2}Experian DataLabs LatAm\\
    \textsuperscript{\rm 3} University of Sao Paulo\\
 
     São Paulo, Brazil\\
    rvilarino@berkeley.edu, rvicente@usp.br

}
\begin{document}

\maketitle

\begin{abstract}
We dissect an experimental credit scoring model developed with real data and demonstrate – without access to protected attributes – how the use of location information introduces racial bias. We analyze the tree gradient boosting model with the aid of a game-theoretic inspired machine learning explainability technique, counterfactual experiments and Brazilian census data. By exposing algorithmic racial bias explaining the trained machine learning model inner mechanisms, this experiment comprises an interesting artifact to aid the endeavor of theoretical understanding of the emergence of racial bias in machine learning systems. Without access to individuals' racial categories, we show how classification parity measures using geographically defined groups could carry information about model racial bias. The experiment testifies to the need for methods and language that do not presuppose access to protected attributes when auditing ML models, the importance of considering regional specifics when addressing racial issues, and the central role of census data in the AI research community. To the best of our knowledge, this is the first documented case of algorithmic racial bias in ML-based credit scoring in Brazil, the country with the second largest Black population in the world.
\end{abstract}

\section{Introduction}

A credit assessment system relies on designing scores that can effectively rank counterparts according to their probability of default. Modern machine learning techniques allow training complex and powerful classifiers at the cost of less explainable outputs \cite{xgboost}.   The standard score development  consists of  a  manual feature engineering stage followed by automatized feature selection, that may be conducted by  ranking features, almost exclusively,  according to their importance for the performance of the classifier.

Here we exemplify the dangers of the common practice by building an experimental credit system based on gradient boosting decision trees, explained by SHAP values \cite{shap}. In particular, we discuss the use of location information, represented in our experiment by the first 3 digits (out of 8 digits) of the Brazilian postal code - we will call this feature CEP-3.  Those 3 digits specify geographical regions larger than individual neighbourhoods and smaller than whole states, depending on the postal granularity of each region. When used for credit scoring, CEP-3 ends up among the most important features performance-wise. However, we show that this feature introduces severe racial biases into the model.

In what follows we analyze the average impact over multiple predictions of the CEP-3 variable as measured by SHAP Values. We found that those averages exhibit strong relation to racial composition differences between different CEP-3. While this result might not sound surprising for some readers, especially those familiar with the US and its practices of Redlining well known by the general public \cite{wikiredlining}. However, discussions around the racial dimensions of territorial occupation are not mainstream in the Brazilian society. Discussions regarding the urban space disparities have been traditionally centered around socioeconomic vulnerability \cite{br-urbano}, following a strong academic and political tradition of dismissing the importance of racial oppression in Brazilian societal disparities \cite{abdias}. The experiment presented here is an important documentation of the interplay between racial dynamics, territorial occupation and machine learning systems for resource allocation in Brazil --- a country central to the history of the Africa diaspora and at the heart of the Atlantic slave trade \cite{abdias}. To the best of our knowledge, this is the first documented case of how algorithmic racial bias may emerge in ML credit scoring built with Brazilian data, the country with the largest Black population outside Africa \cite{abdias}.

The strong relation between CEP-3 impact on model's predictions and territorial racial composition prompted us to ask: how different is using CEP-3 as an input variable from directly using the proportion of not-white people living in someone's neighborhood? The observations we present below very strongly suggest that, at least in the present case, we end up with equivalent models whether we use whiteness proportion or CEP-3. We proceed our presentation analyzing calibration rates --- True Negative Rates, True Positive Rates, False Positive Rates, False Negative Rates --- curves for different thresholds of the model's output in the 5 macro-regions of the country as introduced by the Brazilian Institute of Geography and Statistics (IBGE) in 1970 \cite{fiveregions}. Besides helping us to understand model's behavior countrywide, these curves are an embryonic attempt at procedural evaluation of credit scoring systems in Brazil when protected attributes are not available nor disparities can be linked to any specific input feature.

\section{Model}

We trained a gradient boosting of trees within the XGBoost framework \cite{xgboost} employing a regularized binary logistic objective function. The target was a binary label indicating credit default occurrence in the 1 year period following the input features reference date. Payments over the amount of 100 BRL and delayed by longer than 90 days were accounted as default events. Therefore the model outputs can be interpreted as default probabilities.

\begin{figure}[htbp!]
  \centering
  \includegraphics[width=0.9\columnwidth]{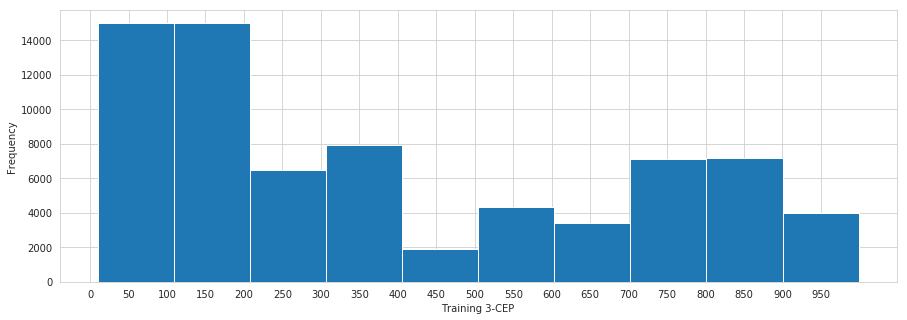}
  \caption{Training data}
  \label{fig:cep-training}
\end{figure}

\begin{figure}[htbp!]
  \centering
  \includegraphics[width=0.9\columnwidth]{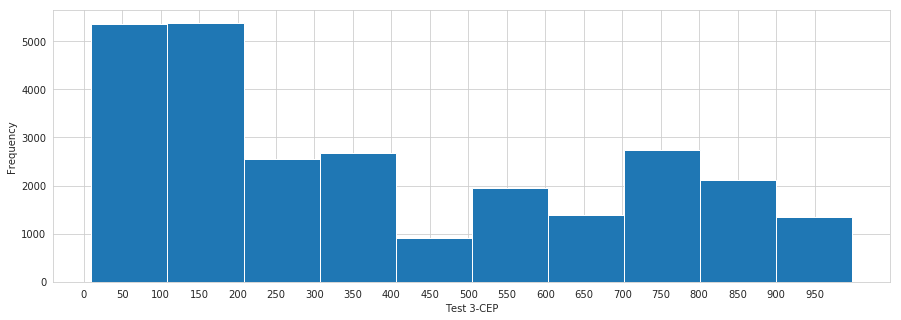}
  \caption{Evaluation data}
  \label{fig:cep-test}
\end{figure}

\begin{figure}[htbp!]
   \centering
   \includegraphics[width=0.9\columnwidth]{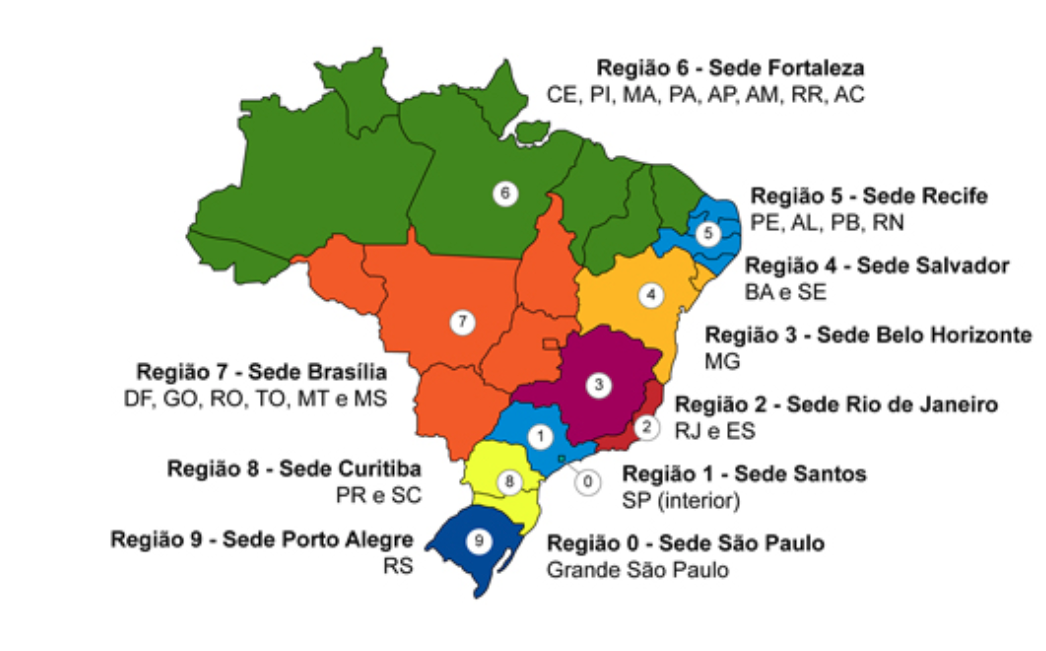}
   \caption{Regions corresponding to first digit of CEP-3.}
   \label{fig:cep-1}
\end{figure}

\begin{figure}[htbp!]
  \centering
  \includegraphics[width=0.9\columnwidth]{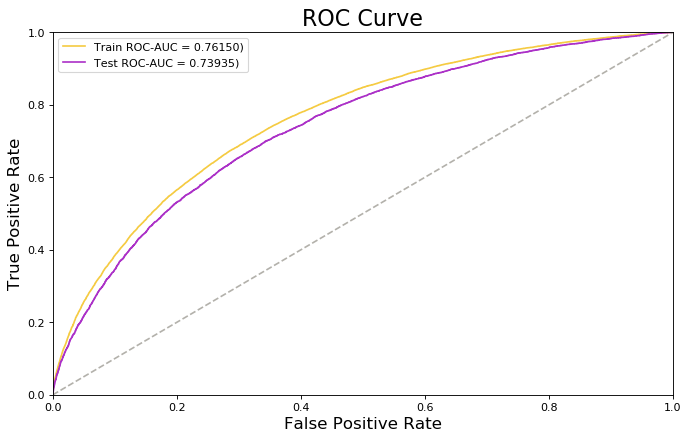}
  \caption{Model's observed ROC curve}
  \label{fig:base-roc}
\end{figure}

\begin{figure}[htbp!]
  \centering
  \includegraphics[width=0.9\columnwidth]{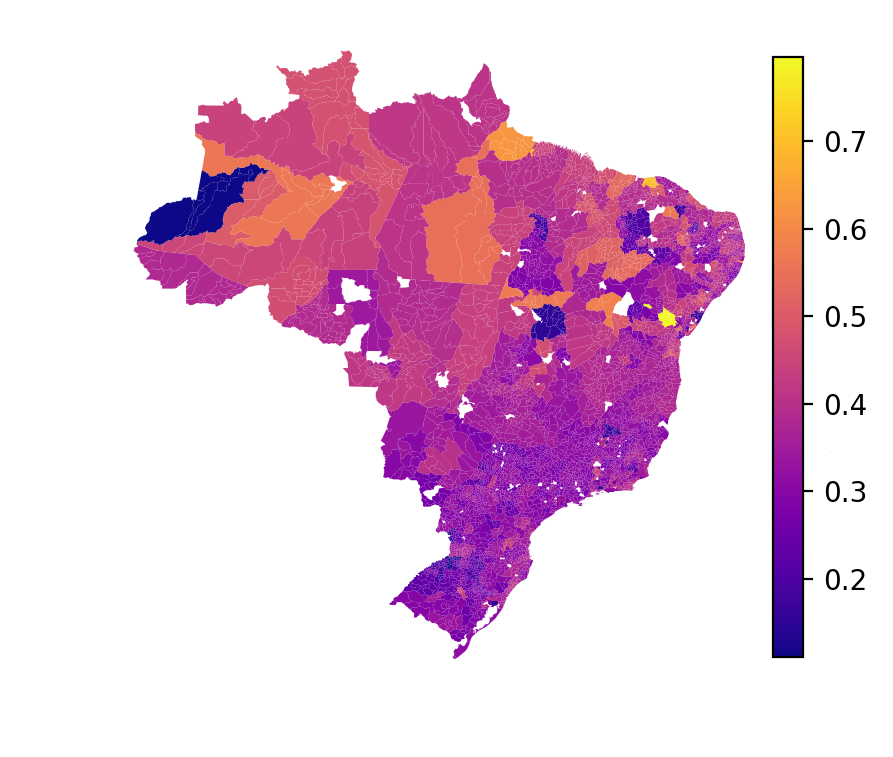}
  \caption{Average default probability output by CEP-3}
  \label{fig:def-prob}
\end{figure}

The data used to build the model consisted of 98 698 examples. Besides the target, each example represented a person through ten features pertaining age, credit and financial history, payment habits, and CEP-3 - each sample was further identified by an anonymized ID and a date referencing the input features. Reference dates ranged from April 1st, 2017 to March 31st, 2018 and entries with reference dates prior to January 1st, 2018 were used for the training dataset, consisting of 72 271 examples. The remaining 26 427 samples were used for performance evaluation. The default rate in the whole data base was 34.5\%, and 34.3\% and 35.2\% for the training and evaluation set respectively. Figures \ref{fig:cep-training} and \ref{fig:cep-test} show the value counts of examples by first digit of the CEP, and figure \ref{fig:cep-1} identifies which regions of the country those digits correspond. The state of São Paulo is overrepresented as the state presents around 20\% of the country's population and accounts for over 40\% of the samples in the development data. This however is common market practice given the importance of the state for the country's economy and consequential increased credit activity --- as baseline, consider that the state was responsible for almost 30\% of the country's GDP in 2019 \cite{pib}. The final model had 250 trees of maximal depth of 3. The model attained a performance of 0.76 and 0.74 of ROC-AUC in train and test, respectively. Figures \ref{fig:base-roc} and \ref{fig:def-prob} shows the model's observed ROC curves along with a geographical plot displaying the average default probability output by the model by CEP-3.


\section{Model Explanations}

SHAP values allow one to assess how the trained model uses the features it is fed with, ascribing a number to the intensity of the impact of each feature in each individual prediction while also indicating if the feature drove the prediction up or down from the average model output. It does so by taking a weighted account of model output, using all possible combinations of presence and absence of features. For an input vector $\mathbf{x}$ and a model $f$, the impact $\phi_{i}$ of the $i$-th feature of $\mathbf{x}$ on the model's output $f(\mathbf{x})$ is given by
\begin{equation}
    \phi_{i}(f, \mathbf{x})
    =
    \frac{1}{|N|}
    \sum_{S \subseteq N\backslash \{i\}}
    \frac{ \left[
    f(\mathbf{x}_{S\cup \{i\}})
    -
    f(\mathbf{x}_{S})
    \right]}{\binom{|N|-1}{|S|}}
\end{equation}
where $N$ is the set of all available features, $S$ represents subsets of features that do not include the $i$ feature and $\mathbf{x}_{S}$ represents a vector containing only the input features in the set $S$. As a proxy to the model's output when it should have access only to a subset of the features, the technique proposes that $f(\mathbf{x}_{S})$ should be evaluated as $E[f(x)|x_{S}]$ \cite{shap2}.

\begin{figure}[htbp!]
    \centering
    \includegraphics[width=0.9\columnwidth]{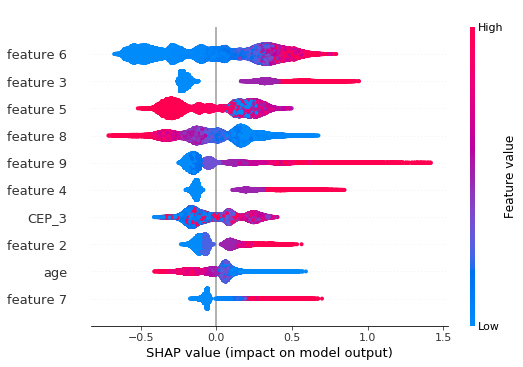}
    \caption{Summary of feature impacts over the dataset.  For example, typically, high values of "age" drive the probability of default down (and the credit score up), while low values of "age" do the opposite.}
    \label{fig:summarycep}
\end{figure}

We analyzed the impact of the features in each prediction on our train and test sets an the results are summarized in figure \ref{fig:summarycep}, which comprises a set of scatter plots, one plot in each line for each input feature. Each individual example is represented by one point in each of the scatter plots. The place of the point along the horizontal axis tells the impact of the respective feature on the model's output of a single prediction. Given a prediction and a feature used in that prediction, a negative SHAP value means that in that prediction, the feature
impacted the model as to lower the model's output probability; while positive SHAP values means that it impacted the prediction as to raise the output probability. The color that each scatter point receives refers to the feature's relative value inside the sample used to build the plot --- relatively high values are pink while relatively low values are blue.

It is interesting to note that almost all features in our credit model behave monotonically, in the sense that either high (low) values of a given feature drive the output probability up (down) or the exact opposite.  In our scenario, this monotonic behavior on SHAP values is probably due to the techniques used in the manual feature engineering and design phase. The methods of feature creation were devised to be optimal for linear models, since these used to be the ones traditionally used within the context of the Brazilian credit market. This could explain why the location variable, CEP-3, is the only one among the most impactful features without a clear monotonic pattern --- since it was not engineered by the same processes as the other ones.

\section{Racial Bias}

\begin{figure}[htbp!]
  \centering
  \includegraphics[width=0.9\columnwidth]{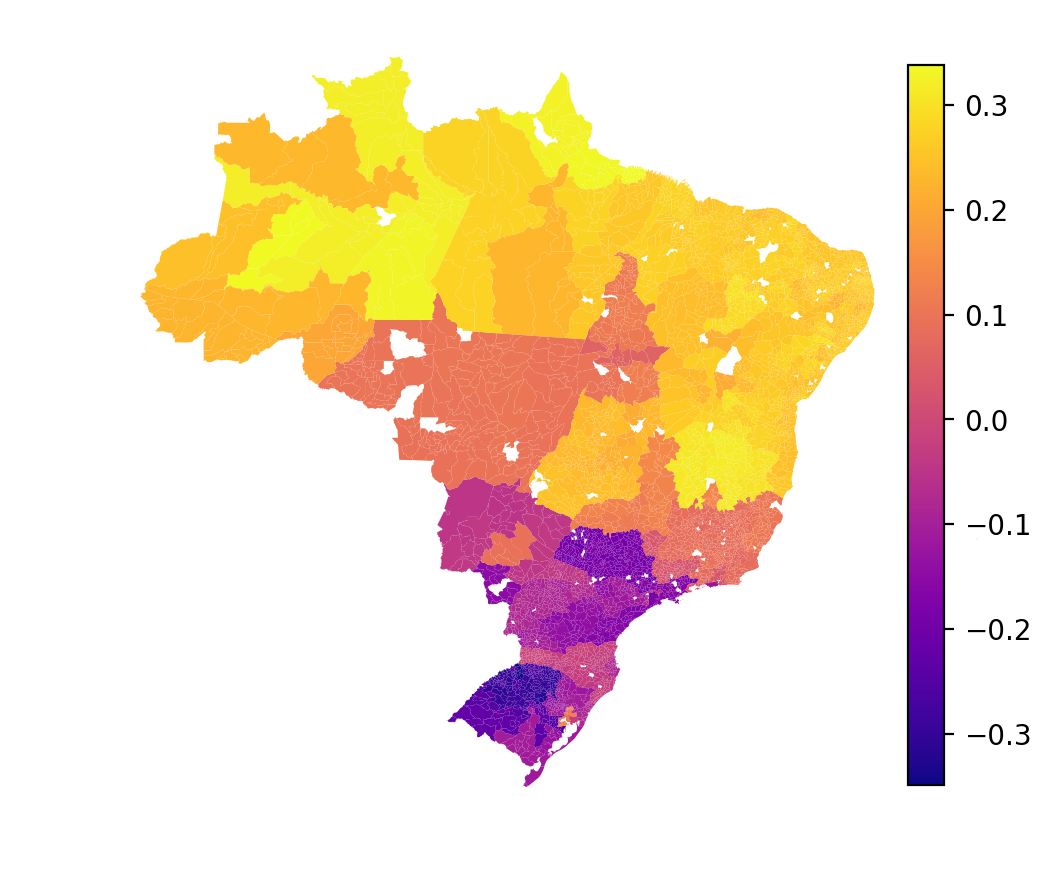}
  \caption{Empirically estimated $E[\phi_{\text{CEP-3}}(f, \mathbf{x})| \text{CEP-3}]$}
  \label{fig:exp-shap}
\end{figure}

\begin{figure}[htbp!]
  \centering
  \includegraphics[width=0.9\columnwidth]{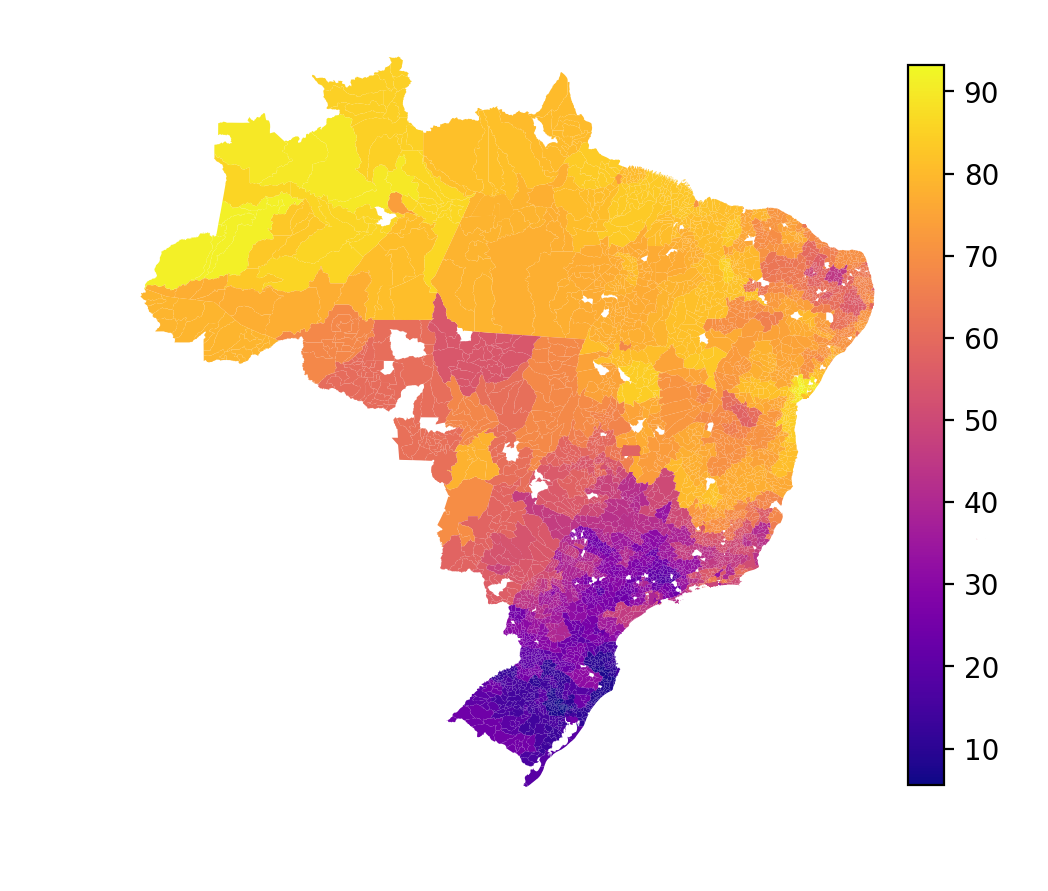}
  \caption{Self-declared not-white proportion by CEP-3}
  \label{fig:per-not-white}
\end{figure}

After being triggered by what appeared to be an strange pattern in the raw values of the SHAP values atributted to CEP-3 in many different predictions, we opted to agregate it to analyze it geographicaly. Let $\phi_{\text{CEP-3}}(f, \mathbf{x})$ denote the SHAP value attributed to the CEP-3 variable in the prediction $f(\mathbf{x})$ for input data $\textbf{x}$. We used our data to empirically estimate $E[\phi_{\text{CEP-3}}(f, \mathbf{x})| \text{CEP-3}]$ and built a geographical plot, as can be seen in figure \ref{fig:exp-shap}. $E[\phi_{\text{CEP-3}}(f, \mathbf{x})| \text{CEP-3}]$ tells us of the average use the tree boost model made of people’s location information. Let us remind ourselves that a negative SHAP value means that the feature impacted the model as to lower the model's output probability, while a positive SHAP value means the feature impacted the prediction as to raise the output default probability. Linking this to figure \ref{fig:exp-shap}, we get that the more purple the region you live in, the better for your credit score.

A deeper meaning of this pattern arises once we compare it to the countrywide geographical racial proportions. We leverage data gathered by the Brazilian Institute of Geography and Statistics (IBGE) in the 2010 census to estimate the proportion of self-declared not-white people by CEP-3 and the result is shown in figure \ref{fig:per-not-white} \cite{IBGE-race}. The resemblance between the two plots is not only uncanny but the data that generated them also exhibits \textbf{83\%} Pearson’s correlation. We refer the reader back to figure \ref{fig:def-prob} to reflect on how by simply looking at geographical distribution of the output probability one would not promptly assume the model had such a strong connection to racial factors --- the correlation between average default probability and proportion of not-whites by CEP-3 is only 48\% in this case. This points to the usefulness and importance of explainability techniques in scrutinizing ML models behavior.

\begin{figure}[htbp!]
    \centering
    \includegraphics[width=0.9\columnwidth]{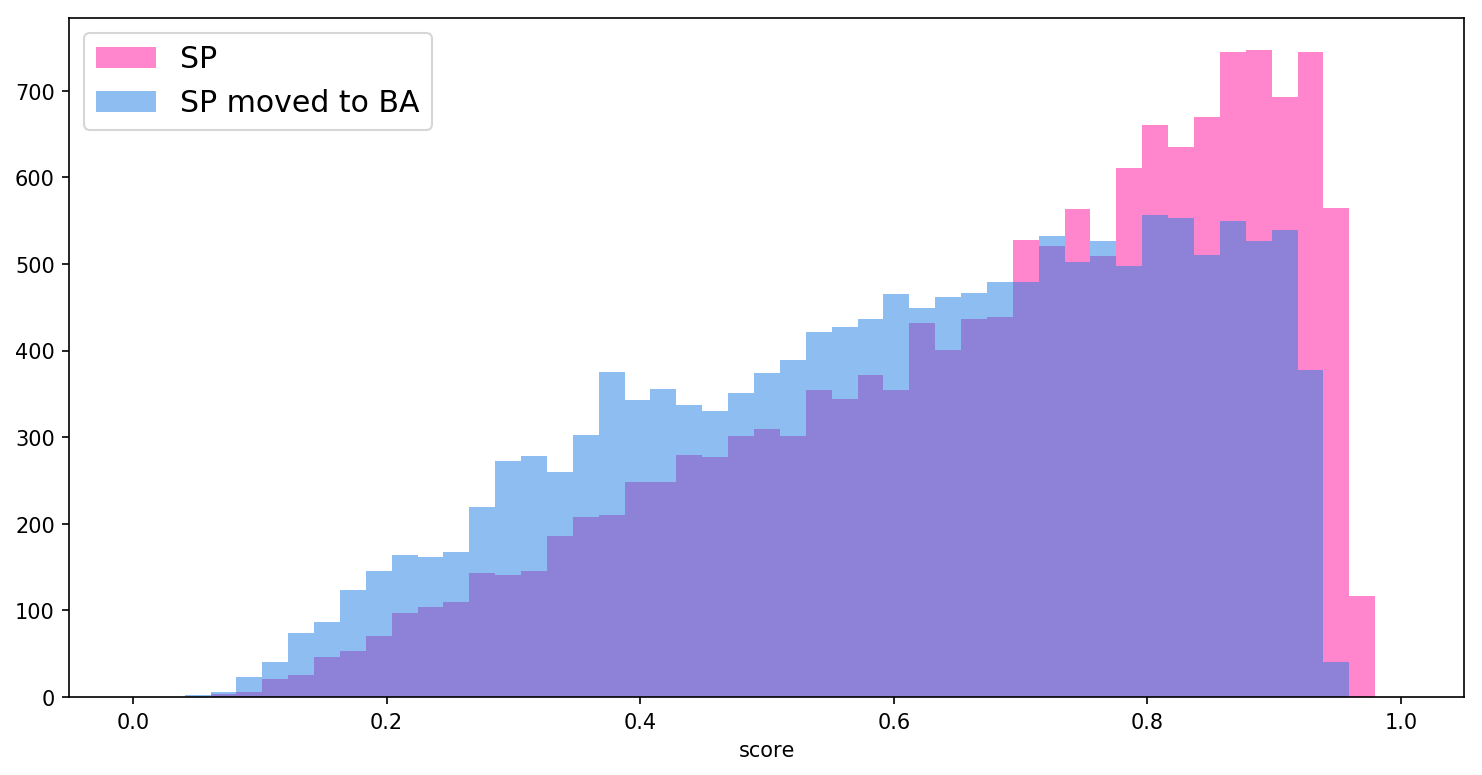}
    \caption{Effect on the credit score distribution when moving from São Paulo (SP) to Bahia (BA). In \textbf{99.8\%} of the cases the credit scores of people moved to Bahia strictly decreases}
    \label{fig:SP-BA}
\end{figure}

One further observation helps us to develop a more concrete sense of the impact of a model like this in real people's lives. We asked ourselves: what would we observe if we moved people away from whiter regions and towards more not-white regions while maintaining all the other attributes --- age, credit and financial history, payment habits --- the same? To get a sense of this effect, we built counterfactual examples by changing the CEP-3 of people living in the state of São Paulo --- where 36\% of the people declare as not-white --- to a CEP-3 codifying somewhere in the state of Bahia, the state with the largest proportion of Black people among Brazilian states, where 78\% of people self-identify as not-white \cite{IBGE-race}. Figure \ref{fig:SP-BA} shows the effect of the simulation on people's credit scores (here in a 0 to 1 scale that is supposed to reflect on credit worthiness). In \textbf{99.8\%} of the cases the credit scores of people moved to Bahia would strictly decrease.

We close this section by highlighting the reasons of our choices to center the present argument around \textit{not-whiteness}. Racial data as gathered by IBGE is based on self-declaration of \textit{color}, as the Brazilian tradition of racialization most frequently refers to skin color rather than to 
ancestry. Another key aspect of Brazilian racialization process is a cultural pressure for people of African ancestry to not identify as Black with \textit{euphemisms} such as \textit{parda/pardo} or \textit{morena/moreno} being pushed to Black people from a very young age as the proper way to refer to themselves \cite{abdias}. A nice and brief summary of these questions in English can be find at \cite{vox-race}. Similarly, there is a even stronger pressure for people of indigenous decent to not identify as such, while the focus on skin color is used for denying the contribution of indigenous people to the Brazilian population composition --- for this frame is used to consider only the people living in traditional indigenous communities as indigenous descendants. All that said, we chose to center the analysis on whiteness because we deem the polarization between white and not-white to better reflect racial tensions in Brazil. We also believe that, by analyzing the data in this frame, our conclusions are less sensitive to the noise due to the cultural pressure to not self-identify as Black or indigenous. In addition, we point to the fact that by centering whiteness in our argument, we put ourselves in contrast to the common sense in Western societies that tends to treat whiteness as a neutral racial category, with harmful effects to technological development \cite{ruha}.

\section{Are CEP-3 and whiteness interchangeable?}

\begin{figure}[htbp!]
  \centering
  \includegraphics[width=0.9\columnwidth]{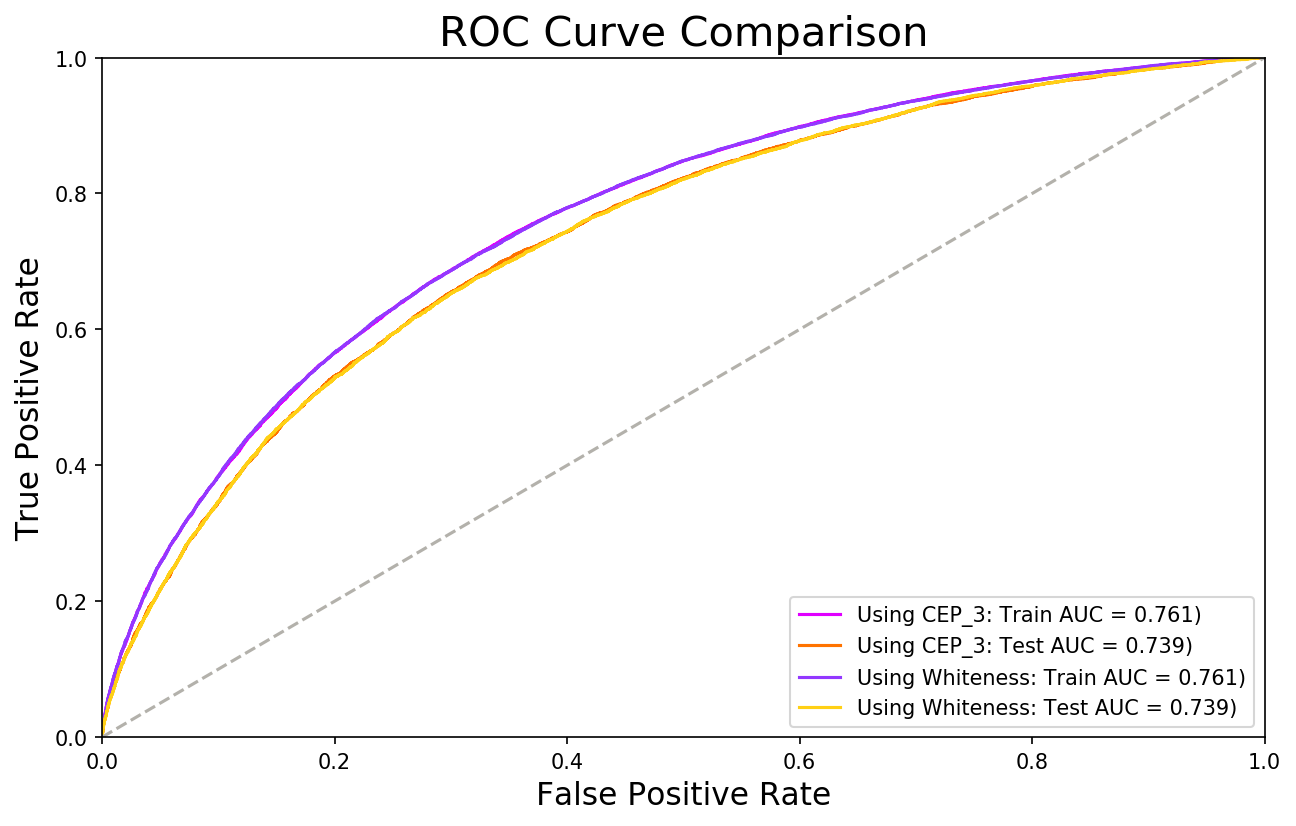}
  \caption{ROC curves.}
  \label{fig:roc-comp}
\end{figure}

\begin{figure}[htbp!]
  \centering
  \includegraphics[width=0.8\columnwidth]{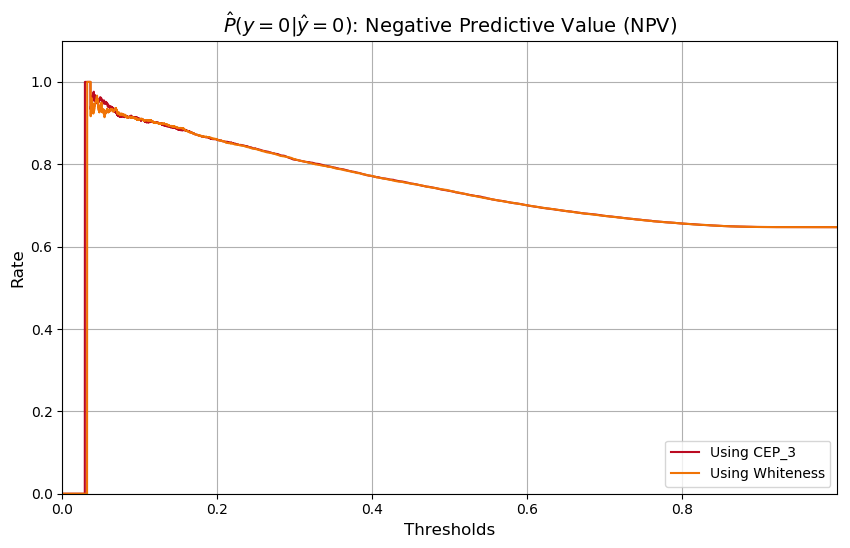}
  \caption{$\hat{p}(y=0|\hat{y}=0)$ - NPV curves.}
  \label{fig:npv-comp}
\end{figure}

\begin{figure}[htbp!]
  \centering
  \includegraphics[width=0.8\columnwidth]{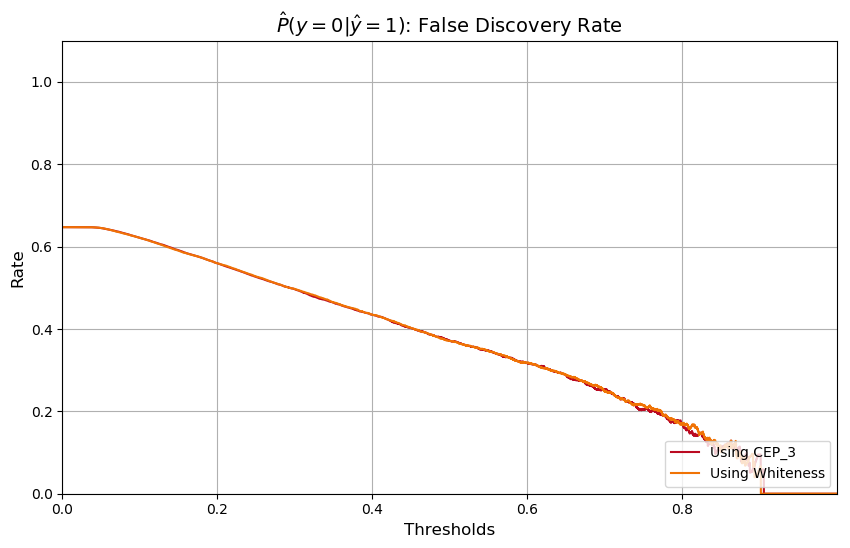}
  \caption{$\hat{p}(y=0|\hat{y}=1)$ - FDR curves.}
  \label{fig:fdr-comp}
\end{figure}

\begin{figure}[htbp!]
  \centering
  \includegraphics[width=0.8\columnwidth]{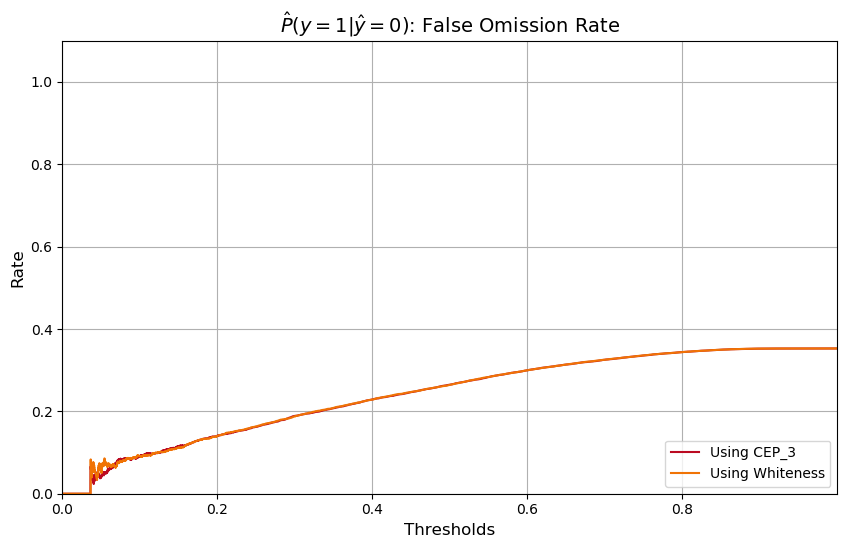}
  \caption{$\hat{p}(y=1|\hat{y}=0)$ - FOR curves.}
  \label{fig:for-comp}
\end{figure}

\begin{figure}[htbp!]
  \centering
  \includegraphics[width=0.8\columnwidth]{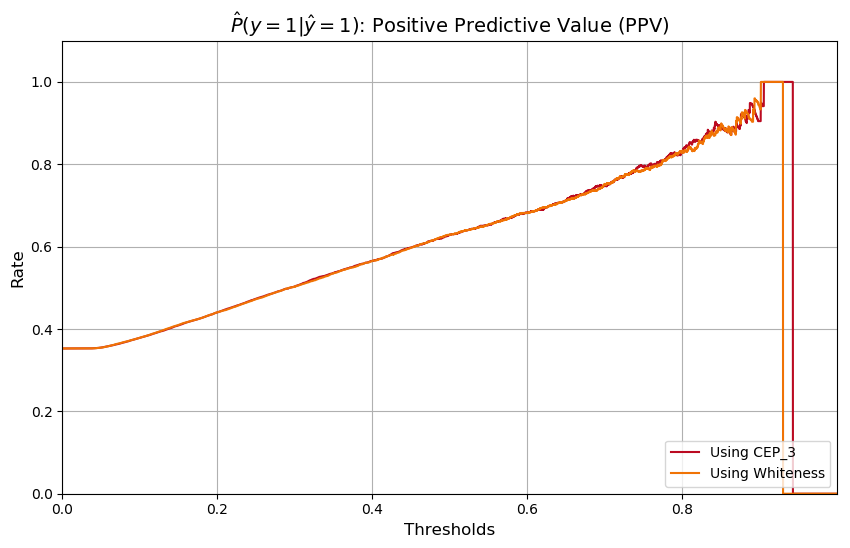}
  \caption{$\hat{p}(y=1|\hat{y}=1)$ - PPV curves.}
  \label{fig:ppv-comp}
\end{figure}

\begin{figure}[htbp!]
  \centering
  \includegraphics[width=0.8\columnwidth]{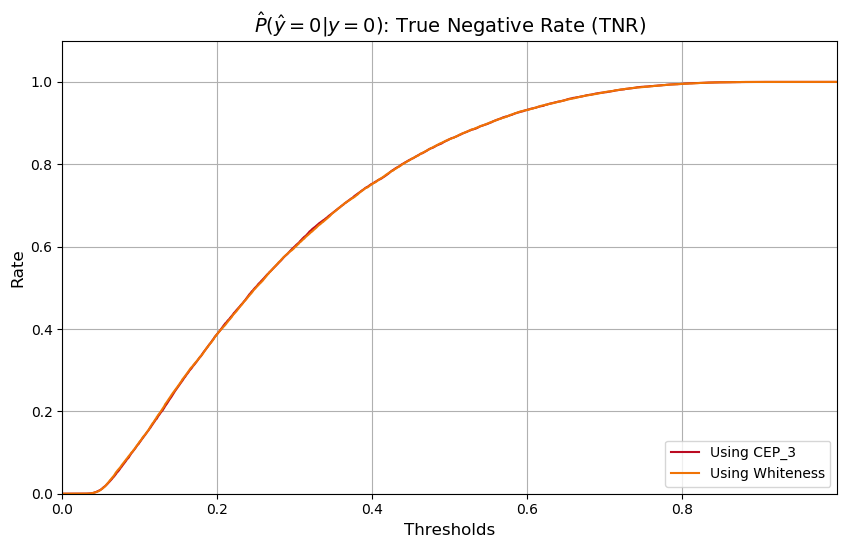}
  \caption{$\hat{p}(\hat{y}=0|y=0)$ - TNR curves.}
  \label{fig:tnr-comp}
\end{figure}

\begin{figure}[htbp!]
  \centering
  \includegraphics[width=0.8\columnwidth]{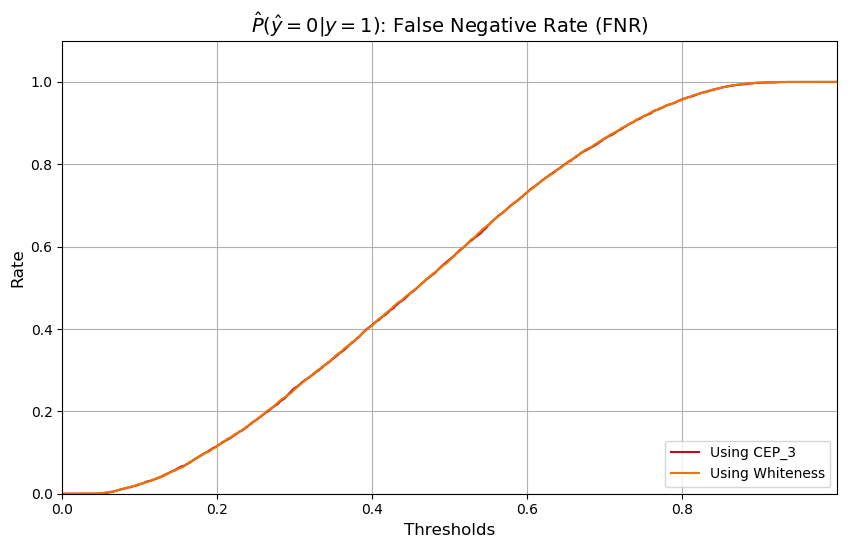}
  \caption{$\hat{p}(\hat{y}=0|y=1)$ - FNR curves.}
  \label{fig:fnr-comp}
\end{figure}

\begin{figure}[htbp!]
  \centering
  \includegraphics[width=0.8\columnwidth]{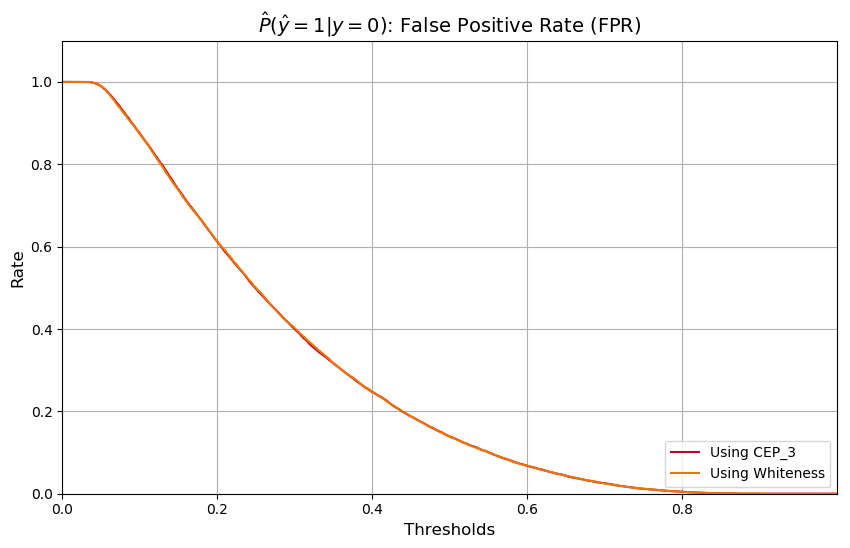}
  \caption{$\hat{p}(\hat{y}=1|y=0)$ - FPR curves.}
  \label{fig:fpr-comp}
\end{figure}

\begin{figure}[htbp!]
  \centering
  \includegraphics[width=0.8\columnwidth]{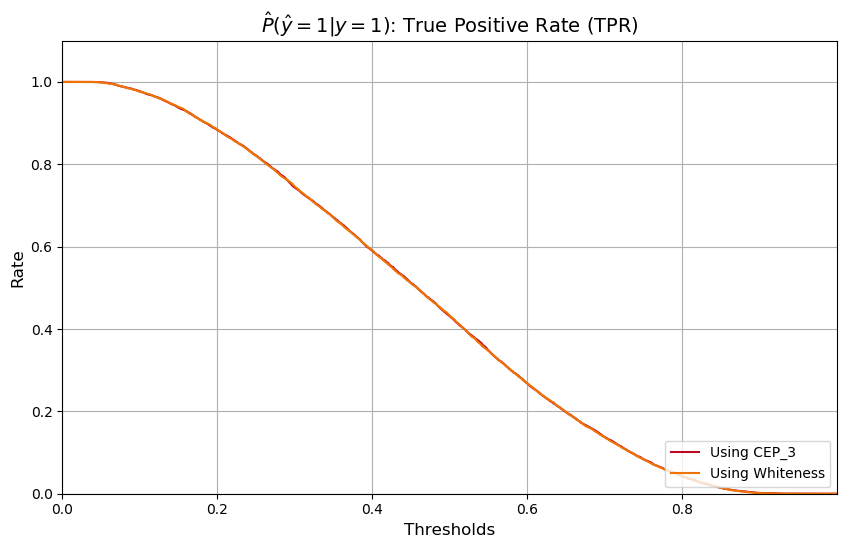}
  \caption{$\hat{p}(\hat{y}=1|y=1)$ - TPR curves.}
  \label{fig:tpr-comp}
\end{figure}

The strong relationship between the impact of CEP-3 on model decisions and the racial distribution raises the question: how would a model that explicitly makes use of neighborhood whiteness instead of the CEP-3 compare to the original one? The plots displayed in figures \ref{fig:roc-comp}-\ref{fig:tpr-comp} suggest we end up with a very similar model, if we must not say an equivalent one. The models exhibit virtually identical ROC curves as well virtually identical NPV, FDR, FOR, PPV, TNR, FNR, FPR and TPR for every threshold.

\begin{figure}[htbp!]
    \centering
    \includegraphics[width=0.9\columnwidth]{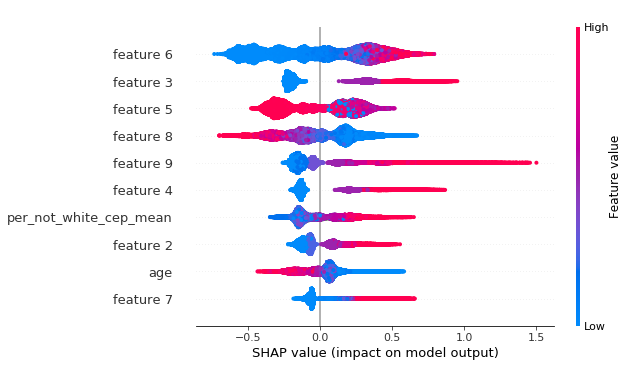}
    \caption{Summary feature impacts over the test dataset for a model using regional racial composition instead of CEP-3 as a feature..}
    \label{fig:summarywhite}
\end{figure}

A summary of the features impact measured by SHAP values in the whiteness based model can be found at figure \ref{fig:summarywhite}. Comparing figures \ref{fig:summarywhite} and \ref{fig:summarycep}, we see that the explanation patterns of the other features did not show any significant change, the new racial variable even takes exactly the place of CEP-3, with the same average importance. A key difference, however, can be seen in the monotonic behavior of the impact of the new feature: the greater the percentage of not-white people living around someone, the higher their impact to increase the model's output default probability ascribed to that someone. While there could be other reasons for the apparent equivalence, the results under this section are certainly telling about the potential harmful effects of using location information to assess credit concession risk.

\section{Possible Generalizations}

\begin{figure}[htbp!]
    \centering
    \includegraphics[width=0.5\columnwidth]{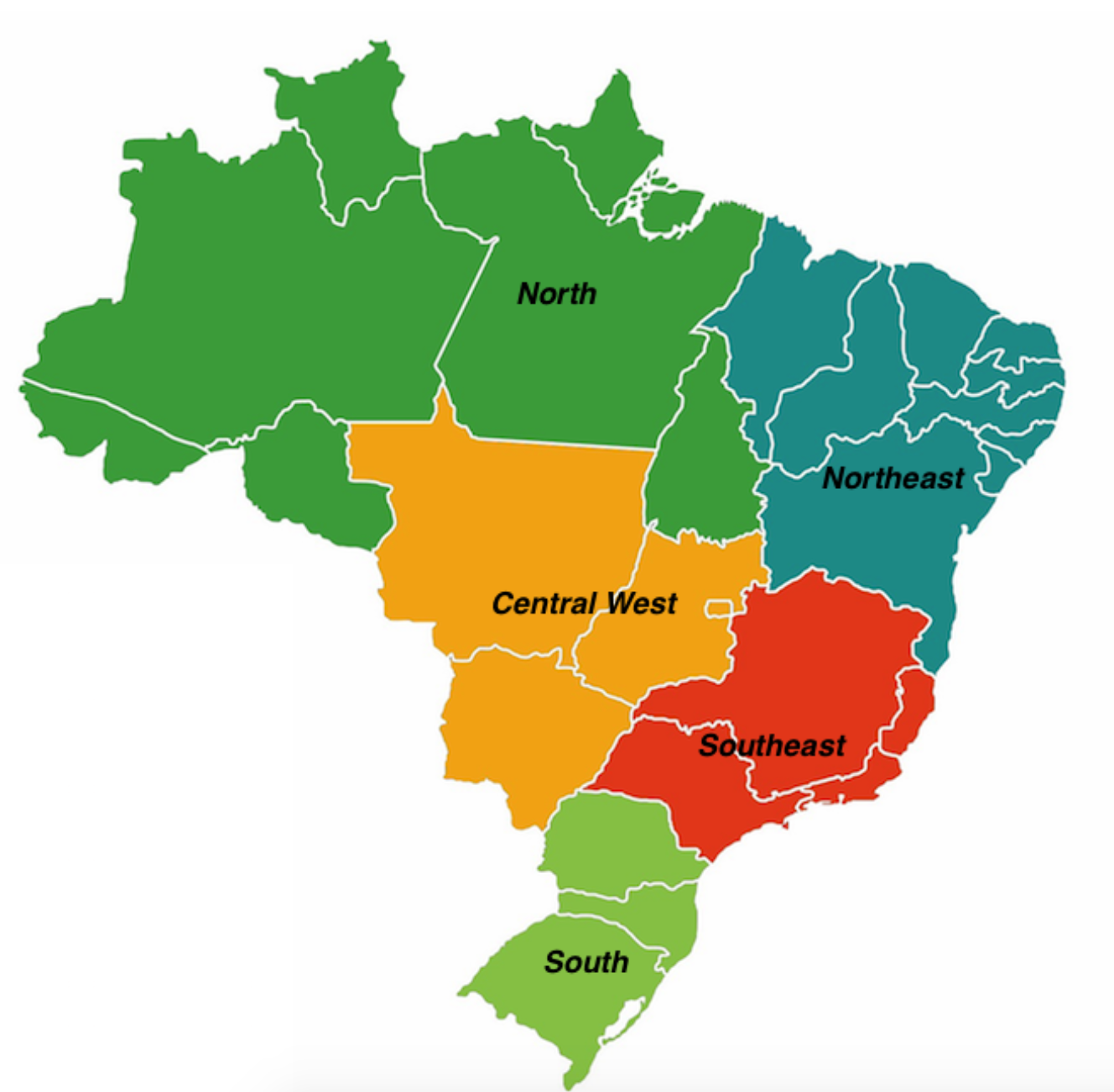}
    \caption{\textbf{Proportion of self-declared not-whites:} North: 76.4 \%; Northeast: 71.2 \%; Central-West: 58.5 \%; Southeast: 43.3 \%; South: 21.5 \%}
    \label{fig:regions}
\end{figure}

We have built a strong argument regarding the racial bias in this model via the use of location information without resourcing to protected attributes. It is crucial to be able to investigate bias and discrimination without referring to protected attributes, especially in a country like Brazil without a strong tradition of collecting racial annotation. Nevertheless, everything discussed up to this point depends on the assessment of the use of CEP-3 by the model. We made an attempt at devising a simple procedure for investigating biases in the Brazilian context without resorting to protected attributes that could be used in more general cases. Figure \ref{fig:regions} shows the division of the country in 5 macro-regions as introduced by IBGE in 1970 \cite{fiveregions}. We resorted to them, since they are very frequently used to reason about countrywide differences in Brazil.

In this embryonic attempt to design some general procedure, we analyzed balance for positive (negative) class between the regions, to understand the probability of being treated by the algorithm to one of the classes conditioned on merit/true outcome. We ask, for example, what is the probability $p(\hat{y}=1|y=0)$ of someone who would not default receive a default prediction from the algorithm? Since credit scores are used in different ways by credit vendors, it makes little sense to investigate this question using binary decisions for an arbitrary threshold. Hence, we built curves that show estimates for empirical conditional probabilities for every possible threshold. This could be turned into a quantitative method by choosing a measure to quantify the disparities between the curves --- e.g. the maximum difference between the values among all thresholds. We decided to now proceed on this way, but we present the curves to point to the kind of approach that could be used in the Brazilian context.

It is worth noting that we opted to analyze probabilities of the kind $p(\hat{y}|y)$ instead of $p(y|\hat{y})$ for two reasons. First, since the later kind of probability is conditioned on model's prediction, curves drawn for different threshold values exhibit discontinuous behavior for high or low threshold values, making it difficult to define threshold intervals on which to base conclusions. Besides this, it is our interpretation that a probability conditioned on the model output tells one about model impact --- since one is assuming the prediction as given, it is equivalent to ask \say{what will happen? \textit{if} the model behaves like this?}. While probabilities conditioned on the true outcome tells one about model treatment --- since one is treating the true outcome as given, it amounts to ask \say{given someone with \textit{this} characteristic, how will the model decide?}. We do not believe model impact assessment should rely exclusively on merit (as happens to be the case in measuring impact using the true outcome as reference), especially in the context of credit scoring and other resource allocation algorithms that can alter the conditions for one to demonstrate merit. Although there are also problems in analyzing model treatment conditioning on true outcome or merit, we believe that it is more fruitful because it helps pointing out more obvious mistreatments. For more on this and other fairness measures limitations, we refer the reader to e.g. \cite{mismeasure}.

\begin{figure}[htbp!]
  \centering
  \includegraphics[width=0.85\columnwidth]{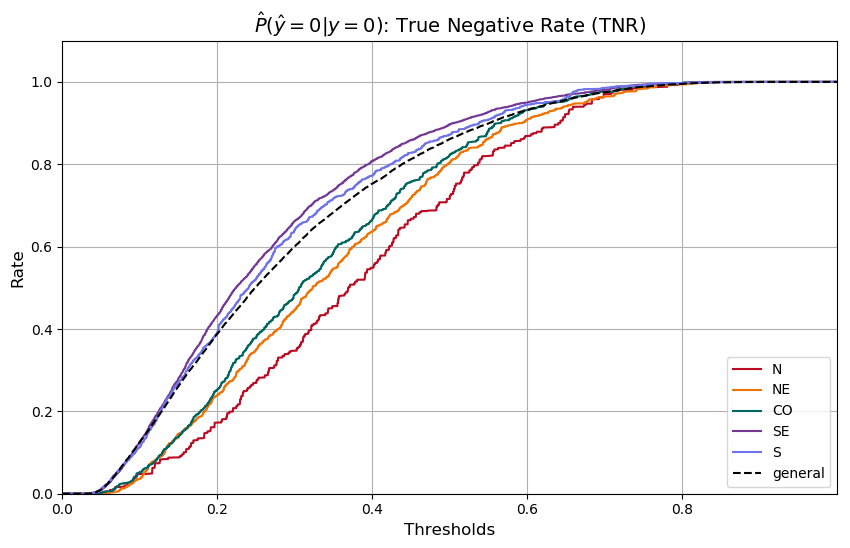}
  \caption{$\hat{p}(\hat{y}=0|y=0)$}
  \label{fig:reg-tnr}
\end{figure}

\begin{figure}[htbp!]
  \centering
  \includegraphics[width=0.85\columnwidth]{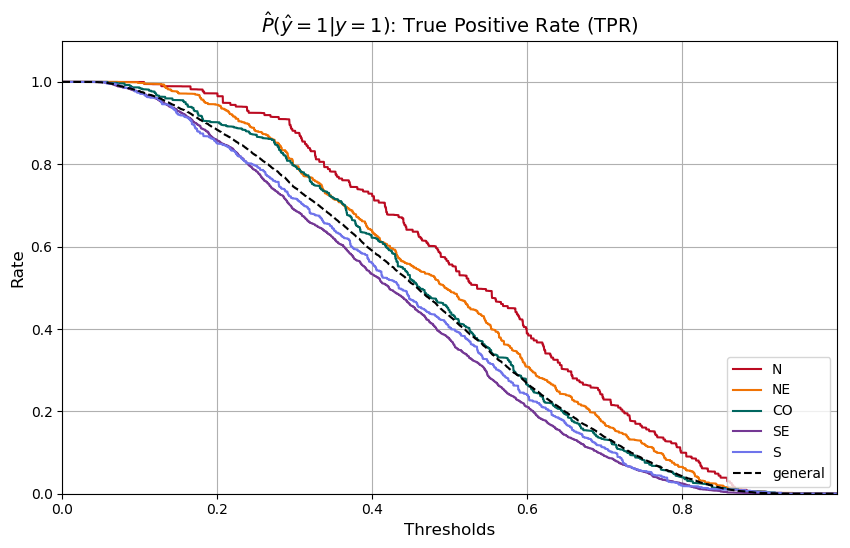}
  \caption{$\hat{p}(\hat{y}=1|y=1)$}
  \label{fig:reg-tpr}
\end{figure}

\begin{figure}[htbp!]
  \centering
  \includegraphics[width=0.85\columnwidth]{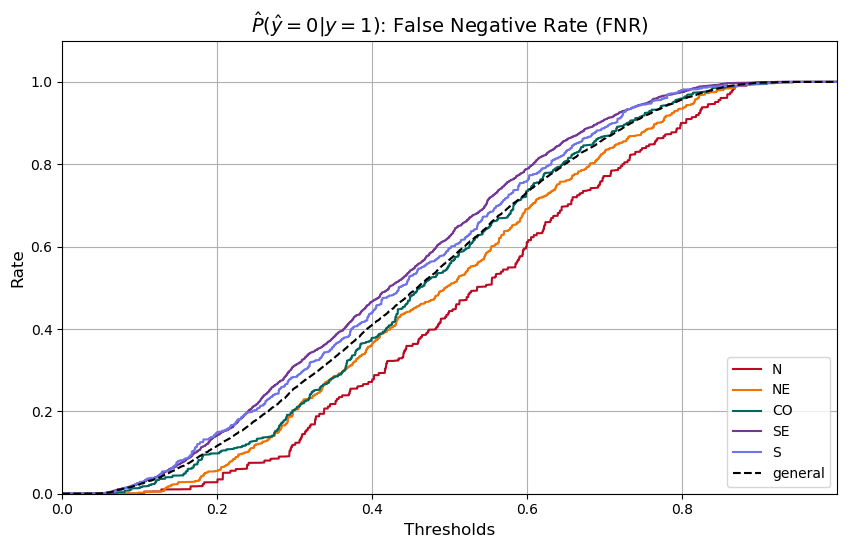}
  \caption{$\hat{p}(\hat{y}=0|y=1)$}
  \label{fig:reg-fnr}
\end{figure}

\begin{figure}[htbp!]
  \centering
  \includegraphics[width=0.85\columnwidth]{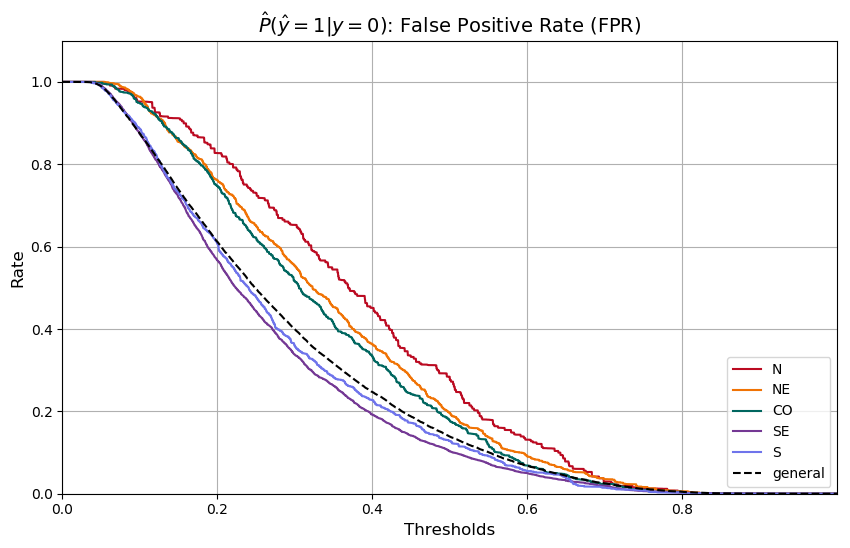}
  \caption{$\hat{p}(\hat{y}=1|y=0)$}
  \label{fig:reg-fpr}
\end{figure}

Figures \ref{fig:reg-tnr} and \ref{fig:reg-tpr} reveal that, \textit{for every threshold}, people in the North, Northeast and Center-West regions --- with more not-white people --- are \textbf{more} likely to be correctly classified for the \textit{non-beneficial} outcome (being considered a default) and \textbf{less} likely to be correctly classified for the \textit{beneficial} outcome than people living in the whiter Southeast and South regions. At the same time, it is more likely for people in the Southeast and South regions to be \textit{misclassified} in a beneficial way (being predicted a non-default when they would be one) and less likely to be \textit{misclassified} in a detrimental way (being classified a default when they would not be one) than people living in the North, Northeast and Center-West regions, as seen in figures \ref{fig:reg-fnr} and \ref{fig:reg-fpr}.

It draws attention that this pattern of misclassifcation (misclassifying in benefit of whiter regions and in detriment of less white regions) resembles that of the famous COMPAS assessment, which would misclassify white and Black defendants in similar fashion \cite{compas}. Given two very different contexts and domains, what makes this result not surprising? We are careful to not overgeneralize from it, for this could be an example of the limitation of this kind of metric for \textit{fairness-like} assessments being raised by some scholars \cite{mismeasure}.

\section{Conclusion}
The experiments described here speak for the importance of explainability techniques for ML in understanding model behavior and reflecting on its consequences. The experiment described here tells us how wider evaluation of model behavior can prompt important questions that might lead to uncovering important biases --- even in the absence of protected attributes. This unprecedented thorough investigation of racial bias in ML systems for credit scoring in Brazil can also contribute to draw attention to the necessity of developing bias tracking framework that does not rely on protected attributes.  The complex and unique history of racial oppression and disparity in Brazil also makes the presented experiment a strong evidence for the need of considering regional specifics when reflecting on racial issues. Finally this article is also a testament for the importance of census data for social justice and AI research.

\section{ Acknowledgments}
We thank Experian DataLab LatAm for supporting this investigation under a collaboration with the Advanced Institute for Artificial Intelligence in São Paulo. We thank the Black in AI organization for fostering important encounters that provided  crucial feedback on our work, with special thanks going to Inioluwa Deborah Raji and Angela Zhou for their contributions in drawing attention to the importance of this work along with crucial feedback on preliminary versions of our writing.

\bibliography{references} 

\end{document}